\documentclass[aps,prd,superscriptaddress,twocolumn]{revtex4-1}
\usepackage{amssymb}
\usepackage{amsmath}
\usepackage{wallpaper}
\usepackage{bm}
\usepackage{multirow}
\usepackage[colorlinks,linkcolor=blue, urlcolor=blue, anchorcolor=blue, citecolor=blue]{hyperref}

\begin{document}

\title{Extended NJL model for baryonic matter and quark matter}

\author{Cheng-Jun Xia}
\email{cjxia@yzu.edu.cn}
\affiliation{Center for Gravitation and Cosmology, College of Physical Science and Technology, Yangzhou University, Yangzhou 225009, China}

\date{\today}

\begin{abstract}
By considering baryons as clusters of three quarks, we extend the Nambu-Jona-Lasinio (NJL) model to describe baryonic matter, quark matter, and their transitions in a unified manner, where the Dirac sea and spontaneous chiral symmetry breaking are considered. In particular, a density-dependent structural function $\alpha_S$ is introduced to modulate the four(six)-point interaction strengths to reproduce the baryon masses in vacuum and medium. Vector interactions are considered with the exchange of $\omega$ and $\rho$ mesons, where the density-dependent coupling constants are fixed by reproducing nuclear matter properties as well as the $\Lambda$-hyperon potential depth in nuclear medium. As density increases, quarks will emerge as quasi-free particles and coexist with baryons. This phase is interpreted as quarkyonic matter, where quarks are restricted to the lowest energy states in the presence of baryons, i.e., quarks are still confined. Similar to the treatment of $\alpha$ clustering in nuclear medium, a Pauli blocking term is added to baryon masses so that baryons eventually become unbound in the presence of a quark Fermi Sea. Then at large enough densities baryons vanish and a Mott transition takes place, we consider such a transition as deconfinement phase transition. Depending on the strengths of Pauli blocking term and quark-vector meson couplings, both first-order and continues phase transitions are observed for quarkyonic, chiral, and deconfinement phase transitions. The corresponding compact star structures are then investigated and confronted with various astrophysical constraints.
\end{abstract}

\maketitle

\section{\label{sec:intro}Introduction}

Cold strongly interacting matter are expected to undergo a deconfinement phase transition at large enough densities, converting hadronic matter into quark matter. Due to the infamous sign problem in lattice QCD and the inability to carry out reliable pertubative QCD calculations, so far it is still unclear at which density does deconfinement phase transition take place or whether it is of first order or a smooth crossover~\cite{Fukushima2005_PRD71-034002, Voskresensky2023_PPNP130-104030}. Such a situation may be improved in the era of multi-messenger astronomy, where recent Bayesian analysis~\cite{Xie2021_PRC103-035802, Annala2023_NC14-8451, Han2023_SB68-913, Pang2024_PRC109-025807} and binary neutron star merger simulations~\cite{Bauswein2019_PRL122-061102, Huang2022_PRL129-181101, Fujimoto2023_PRL130-91404} have shed light on identifying a possible deconfinement phase transition in massive compact stars.

For a first-order deconfinement phase transition~\cite{Dexheimer2010_PRC81-045201}, the properties of hadronic matter and quark matter are distinctively different. A quark-hadron mixed phase is thus expected in hybrid stars, which may exhibit various geometrical structures based on the surface tension of quark-hadron interface and consequently impact hybrid star structures~\cite{Heiselberg1993_PRL70-1355, Glendenning2001_PR342-393, Voskresensky2002_PLB541-93, Tatsumi2003_NPA718-359, Voskresensky2003_NPA723-291, Endo2005_NPA749-333, Maruyama2007_PRD76-123015, Peng2008_PRC77-065807, Yasutake2014_PRC89-065803, Maslov2019_PRC100-025802, Xia2019_PRD99-103017, Xia2020_PRD102-023031}. Alternatively, the hadron-quark crossover at finite densities was proposed, which predicts a stiff equation of state (EOS) to support massive hybrid stars~\cite{Baym1979_PA96-131, Celik1980_PLB97-128, Schaefer1999_PRL82-3956, Fukushim2004_PLB591-277, Hatsuda2006_PRL97-122001, Maeda2009_PRL103-085301, Masuda2013_ApJ764-12, Masuda2013_PTEP2013-073D01, Zhao2015_PRD92-054012, Kojo2015_PRD91-045003, Masuda2016_EPJA52-65, Whittenbury2016_PRC93-035807, Bai2018_PRD97-023018, Baym2019_ApJ885-42}. As demonstrated by Fukushima and Kojo~\cite{Fukushima2016_ApJ817-180}, the crossover from hadronic matter to quark matter can be bridged by quarkyonic matter, which is comprised of ``a quark Fermi Sea'' and ``a baryonic Fermi surface''~\cite{McLerran2009_NPA830-709c, McLerran2007_NPA796-83, McLerran2009_NPA824-86}.

At densities below and around the nuclear saturation density $n_{0}\approx 0.16$ fm${}^{-3}$, it was shown that treating nucleons as the basic building blocks gives satisfactory description for finite nuclei and nuclear matter, e.g., relativistic mean field (RMF) models~\cite{Meng2016_RDFNS}, nonrelativistic density functional methods~\cite{Dutra2012_PRC85_035201}, and microscopic many-body theories~\cite{Hebeler2021_PR890-1}. However, in nuclear medium the effects of quark degrees of freedom will emerge, such as the EMC effect with the nuclear structure functions differ from those of free nucleons~\cite{Aubert1983_PLB123-275}. By treating nucleons as three quarks confined in bags with the interaction mediated by mesons, the quark-meson-coupling (QMC) model was proposed, where the EMC effect in the valence region could be understood~\cite{Guichon2018_PPNP100-262}.

When density becomes larger, it is nevertheless insufficient to treat baryons as independent particles, where the number of exchanged quarks between baryons increase and baryons may overlap with each other~\cite{Fukushima2016_ApJ817-180, McLerran2009_NPA830-709c, Park2021_PRD104-094024, Park2022_PRD105-114034}. It was shown that quark wave functions simultaneously exist in different baryons to support the attractive interactions in the intermediate range~\cite{Chen2007_PRC76-014001}, while the short range repulsion is provided by the Pauli principle~\cite{Oka1980_PLB90-41}. In such cases, to unveil the microscopic dynamics of baryonic matter, quark matter, and their transitions, a unified description based on quark degrees of freedom is essential, where extensive efforts were made in the past decades.
For example, Horowitz and Piekarewicz investigated the hadron-quark transition based on a nonrelativistic one dimensional string-flip model~\cite{Horowitz1991_PRC44-2753}. A microscopic approach of color molecular dynamics was proposed by Maruyama and Hatsuda~\cite{Maruyama2000_PRC61-062201}, where quarks are clusterized into baryons at low density and a deconfined quark matter is formed at high density via crossover~\cite{Yasutake2024_PRD109-043056}. By treating nucleon as a quark-diquark state using the Nambu-Jona-Lasinio (NJL) model, it was shown that the effects of confinement leads to a scalar polarizability of the nucleon and a less attractive effective interaction between nucleons, helping to achieve saturation of the nuclear matter ground state~\cite{Bentz2001_NPA696-138}. The hadronic SU(3) nonlinear $\sigma$ model was extended to quark degrees of freedom, suggesting that the deconfinement phase transition is of first order~\cite{Dexheimer2010_PRC81-045201}. A direct modeling of quarkyonic matter was recently proposed, where the pressure and sound velocity increase rapidly with density~\cite{McLerran2019_PRL122-122701, Margueron2021_PRC104-055803}. By synthesizing the Walecka model and quark-meson model, a complete field model was developed~\cite{Cao2020_JHEP10-168, Cao2022_PRD105-114020}, which well describes the chiral symmetry restoration in quarkyonic matter.

In our previous study, by combining RMF models and equivparticle models with density-dependent quark masses, we have investigated nuclear matter, quarkyonic matter, and quark matter in a unified manner~\cite{Xia2018_JPSCP20-011010, Xia2023_PRD108-054013}. Nevertheless, the dynamic chiral symmetry restoration was not addressed, and the contribution of Dirac sea was neglected. To better describe strongly interacting matter at vast density ranges, it is favorable to start from quark degrees of freedom directly and consider baryons as clusters made of three valence quarks. As density increases, similar to the melting of light clusters in nuclear medium~\cite{Typel2010_PRC81-015803}, the deconfinement phase transition can be viewed as a Mott transition of quark clusters~\cite{Bastian2018_Universe4-67}.
In this work, we thus extend the NJL model to describe baryonic matter, quark matter, and their transitions in a unified manner, where the Dirac sea and spontaneous chiral symmetry breaking are accounted for. To reproduce the baryon masses in vacuum and medium, the four(six)-point interaction strengths in NJL model are modulate by a density-dependent structural function. The repulsive interaction is treated with the exchange of $\omega$ and $\rho$ mesons, where density-dependent couplings are fixed by reproducing nuclear matter properties as well as the $\Lambda$-hyperon potential depth in nuclear medium.

The paper is organized as follows. In Section~\ref{sec:the}, we present the theoretical framework for the extended NJL model with the model parameters fixed based on various constraints. Several types of phase transitions and the properties of dense stellar matter are then examined, where the corresponding compact star structures are confronted with various astrophysical constraints in Section~\ref{sec:res}. We draw our conclusion in Section~\ref{sec:con}.

\section{\label{sec:the}Theoretical framework}
\subsection{\label{sec:the_Lagrangian} Lagrangian density}
In the mean-field approximation, the Lagrangian density of an extended SU(3) NJL model is given by
\begin{eqnarray}
\mathcal{L} &=& \sum_{i} \bar{\psi}_i \left(  i \gamma_\mu D^\mu_i - M_i \right)\psi_{i} - \frac{1}{4} \omega_{\mu\nu}\omega^{\mu\nu}  + \frac{1}{2}m_\omega^2 \omega^2   \nonumber\\
 &&\mbox{}    + \frac{1}{2}\sum_{i=u,d,s} \left[
\partial_\mu \sigma_i \partial^\mu \sigma_i - m_\sigma^2 \sigma_i^2 \right] - \frac{1}{4} \vec{\rho}_{\mu\nu}\cdot\vec{\rho}^{\mu\nu}  \nonumber \\
 &&\mbox{}  + \frac{1}{2}m_\rho^2 \rho^2 +4K  \bar{n}_{u}^{s}\bar{n}_{d}^{s}\bar{n}_{s}^{s}.  \label{eq:Lagrangian}
\end{eqnarray}
Here $\psi_{i}$ represents the Dirac spinor for different fermions $i$, where in this work we have included baryons ($p$, $n$, $\Lambda$), quarks ($u$, $d$, $s$) and leptons ($\mu$, $e$). By replacing the quark condensations with $\sigma_{u,d,s}$ fields, the four-point interaction becomes nonlocal, while the six-point (`t Hooft) interaction remains local. In principle, the diquark coupling terms can also be included to treat pairings among baryons and quarks, which should be explored in our future study. Vector meson fields are also introduced to account for the repulsive interactions, where the field tensors are
\begin{equation}
\omega_{\mu\nu} = \partial_\mu \omega_\nu - \partial_\nu \omega_\mu,  \ \
\vec{\rho}_{\mu\nu} = \partial_\mu \vec{\rho}_\nu - \partial_\nu \vec{\rho}_\mu.
\end{equation}
The covariant derivative in Eq.~(\ref{eq:Lagrangian}) takes the form
\begin{equation}
i D^\mu_i = i \partial^\mu -  f_i g_{\omega} \sum_{q=u,d,s} N^q_i \omega^\mu  - f_i g_{\rho}  \vec{\tau}_i\cdot\vec{\rho}^\mu,
\end{equation}
where $N^q_i$ is the number of valence quarks $q$ in particle $i$ and $\vec{\tau}_i$ the isospin. The factor $f_i$ modulates the coupling strengths between particle $i$ and vector mesons, while $g_{\omega}$ and $g_{\rho}$ are fixed according to nuclear matter properties with $f_p=f_n =1$. In this work, we consider baryons as clusters of quarks with their effective masses given by
\begin{equation}
 M_i = \sum_{q=u,d,s} N^q_i \left[m_{q0} + \alpha_S(M_{q}-m_{q0})\right]  + B n_\mathrm{b}^Q \label{eq:Bmass}
\end{equation}
and quark masses by
\begin{equation}
 M_{q} = m_{q0} - g_{\sigma} \sigma_q + 2 K \frac{\bar{n}_{u}^{s}\bar{n}_{d}^{s}\bar{n}_{s}^{s}}{\bar{n}_q^{s}} \label{eq:qmass}
\end{equation}
with the effective quark scalar density
\begin{equation}
  \bar{n}_q^s =  n_{q}^{s}+\alpha_S \sum_{i=p,n,\Lambda} N^q_i  n_{i}^{s},
\end{equation}
where $n_{i}^{s}$ is fixed by Eq.~(\ref{eq:ns}). A density dependent structural function $\alpha_S$ is introduced for baryons, which mimics the dampened interaction strength as chiral condensates diminish within baryons~\cite{Bentz2001_NPA696-138, Reinhardt2012_PRD85_074029, Xia2014_CPL31-41101}. The last term in Eq.~(\ref{eq:Bmass}) accounts for the effects of Pauli blocking and interactions between quarks and baryons with $n_\mathrm{b}^Q=(n_u+n_d+n_s)/3$ being the baryon number density of quarks~\cite{Xia2018_JPSCP20-011010, Xia2023_PRD108-054013}, which resembles the treatments of $\alpha$ clustering inside nuclear matter~\cite{Roepke2014_PRC90-034304, Xu2016_PRC93-011306}. The total baryon number density is then
\begin{equation}
  n_\mathrm{b}=n_p+n_n+n_\Lambda+n_\mathrm{b}^Q,
\end{equation}
where the number density of particle $i$ is determined by Eq.~(\ref{eq:ni}). The masses of leptons remain constant with $M_e=0.511~$MeV and $M_{\mu}=105.66~$MeV~\cite{Olive2014_CPC38-090001}.

Due to time-reversal symmetry and charge conservation, the boson fields take mean values with only the time component and the $3$rd component in the isospin space. We then define $\omega\equiv \omega_0$, $\rho\equiv \rho_{0,3}$, and $\tau_i\equiv \tau_{i,3}$ for simplicity. Based on the Lagrangian density in Eq.~(\ref{eq:Lagrangian}), the meson fields of uniform dense matter are determined by
\begin{eqnarray}
m_\sigma^2 \sigma_q &=&  g_{\sigma}\bar{n}_q^s \label{eq:KG_vsigma} \\ 
m_\omega^2 \omega &=& g_{\omega} \sum_i f_i \sum_{q=u,d,s} N^q_i  n_i, \label{eq:KG_omega}\\
m_\rho^2 \rho     &=& g_{\rho} \sum_i f_i \tau_i n_i. \label{eq:KG_rho}
\end{eqnarray}
At vanishing temperatures, the number density and scalar density of fermion $i$ in uniform dense matter are
\begin{eqnarray}
n_{i}&=&\langle \bar{\psi}_{i} \gamma^{0} \psi_{i} \rangle = \frac{g_{i}\nu_{i}^{3}}{6\pi^{2}}, \label{eq:ni}\\
n_{i}^{s}&=&\langle \bar{\psi}_{i} \psi_{i} \rangle = \frac{g_i M_{i}^{3}}{4\pi^{2}}\left[x_{i}\sqrt{x_{i}^{2}+1}-\mathrm{arcsh}(x_{i}) \right.  \nonumber \\
&&   \left. - y_{i}\sqrt{y_{i}^{2}+1}+\mathrm{arcsh}(y_{i})\right]. \label{eq:ns}
\end{eqnarray}
Here we take $x_{i} \equiv \nu_{i}/M_{i}$ with $\nu_{i}$ being the Fermi momentum, $y_{i} \equiv \Lambda/M_{i}$ with $\Lambda$ being the 3-momentum cut-off to regularize the vacuum part of quarks ($\Lambda=0$ for baryons and leptons), and the degeneracy factors $g_{n,p,\Lambda}=g_{e,\mu}=2$ and $g_{u,d,s}=6$. Note that the baryonic Dirac sea does not exist since quarks no longer form clusters in the Dirac sea due to Pauli blocking. Meanwhile, it is worth mentioning that in the quarkyonic phase with baryons and quarks coexisting in a same volume, only baryons can be exited to higher energy states while quarks are restricted to the lowest energy states as they are still confined.

The energy density of uniform dense matter is fixed by
\begin{eqnarray}
E&=& \sum_{i} \frac {g_i{M_i}^4}{16\pi^{2}} \left[x_i(2x_i^2+1)\sqrt{x_i^2+1}-\mathrm{arcsh}(x_i) \right] \nonumber \\
&&  -\sum_{i=u,d,s} \frac {g_i{M_i}^4}{16\pi^{2}} \left[y_i(2y_i^2+1)\sqrt{y_i^2+1}-\mathrm{arcsh}(y_i) \right] \nonumber \\
&&   + \frac{1}{2}\left[ m_\omega^2 \omega^2  + m_\rho^2 \rho^2 + \sum_{i=u,d,s}m_\sigma^2 \sigma_i^2 \right] -4K \bar{n}_{u}^{s}\bar{n}_{d}^{s}\bar{n}_{s}^{s} \nonumber \\
&&      - {E}_0. \label{eq:ener}
\end{eqnarray}
Here a constant $E_0$ is introduced to ensure $E = 0$ in the vacuum. Note that the energy contributions of mesons are obtained with $m_\phi^2 \phi^2/2$ ($\phi=\sigma,\omega,\rho$), while in practice we substitute Eqs.~(\ref{eq:KG_vsigma}-\ref{eq:KG_rho}) into Eq.~(\ref{eq:ener}) so that the meson fields $\phi$ are not calculated explicitly and the meson masses only appear in combination with the couplings as $g_\phi^2/m_\phi^2$. The chemical potentials of baryon $b$, quark $q$, and lepton $l$ are fixed by
\begin{eqnarray}
 \mu_b &=& \sqrt{\nu_b^{2}+M_b^{2}} + f_b(3  g_{\omega} \omega +  g_{\rho} \tau_{b}  \rho)   +\Sigma_b^\mathrm{R}, \label{eq:chem_b}\\
 \mu_{q} &=&\sqrt{\nu_q^{2}+M_{q}^2}   + f_q(g_{\omega} \omega +g_{\rho} \tau_q \rho) + \Sigma_{q}^\mathrm{R}, \label{eq:chem_q}\\
 \mu_{l} &=&\sqrt{\nu_l^{2}+M_{l}^2}, \label{eq:chem_l}
\end{eqnarray}
with the ``rearrangement'' terms given by
\begin{eqnarray}
 \Sigma_b^\mathrm{R}&=& \sum_{i} f_i\left( \omega n_{i}
 \sum_{q=u,d,s} N^q_i \frac{\mbox{d} g_{\omega}}{\mbox{d} n_\mathrm{b}} +  \rho \tau_{i} n_{i} \frac{\mbox{d} g_{\rho}}{\mbox{d} n_\mathrm{b}}\right)
  \nonumber \\
 && +\sum_{i=n,p,\Lambda} \left[ \frac{\mbox{d}  \alpha_S}{\mbox{d} n_\mathrm{b}}\sum_{q=u,d,s} N^q_i(M_{q}-m_{q0}) \right] n_{i}^s, \label{eq:Sigma_b} \\
 \Sigma_{q}^\mathrm{R}&=& \frac{1}{3} B \sum_{i=n,p,\Lambda}n_{i}^s + \frac{1}{3}\Sigma_b^\mathrm{R}. \label{eq:Sigma_Q}
\end{eqnarray}
Then the pressure $P$ is obtained with
\begin{equation}
 P=\sum_{i} \mu_{i} n_{i}-E. \label{eq:pressure}
\end{equation}

\subsection{\label{sec:the_Param} Model parameters}

In this work we adopt the RKH parameter set of NJL model, i.e., $\Lambda = 602.3$ MeV, $m_{u0}= m_{d0} = 5.5$ MeV,  $m_{s0}=  140.7$ MeV, $G_S = g_{\sigma}^2/4m_{\sigma}^2= 1.835/\Lambda^2$, and $K = 12.36/\Lambda^5$~\cite{Rehberg1996_PRC53-410}. This parameter set well reproduces the meson properties, i.e., $m_\pi=135.0$ MeV, $m_K=497.7$ MeV, $m_{\eta'}=957.8$ MeV, $m_{\eta}=514.$ MeV, and $f_\pi=92.4$ MeV~\cite{Rehberg1996_PRC53-410}. The corresponding vacuum quark condensates are $\bar{n}_{u}^s=\bar{n}_{d}^s = -1.843\ \mathrm{fm}^{-3}$ and $\bar{n}_{s}^s = -2.227\ \mathrm{fm}^{-3}$, indicating the quark masses $M_{u}=M_{d} = 367.6\ \mathrm{MeV}$ and $M_{s}= 549.5\ \mathrm{MeV}$.

\begin{figure}[!ht]
  \centering
  \includegraphics[width=0.95\linewidth]{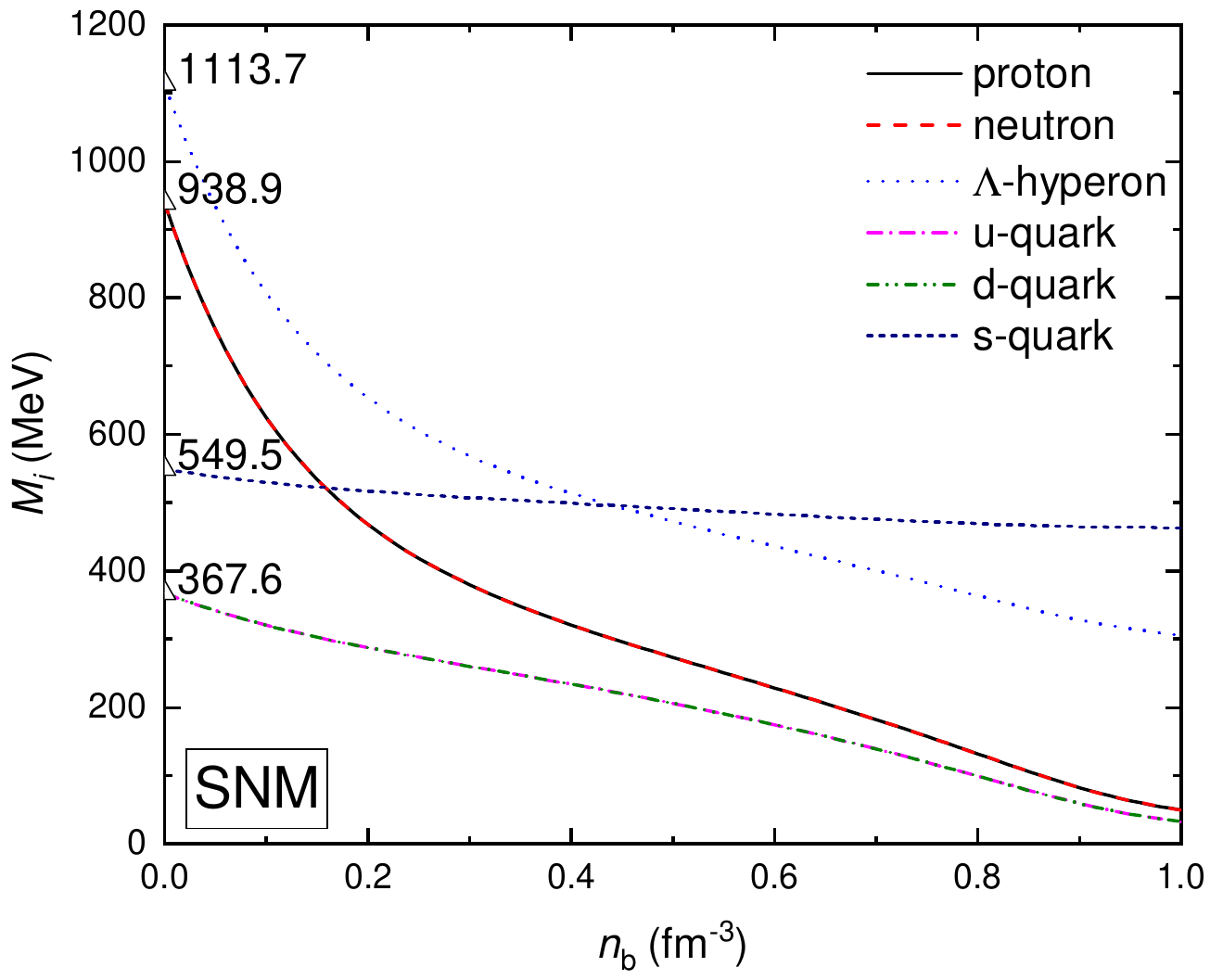}
  \caption{\label{Fig:Mi_sym} Masses of nucleons, $\Lambda$-hyperons, and quarks in symmetric nuclear matter (SNM) as functions of the total baryon number density $n_\mathrm{b}$, which are determined by adopting the density dependent structural function~(\ref{eq:alphaS}).}
\end{figure}

To reproduce the masses of nucleons and $\Lambda$-hyperons in vacuum, we take $\alpha_S = 0.849$ at $n_\mathrm{b}=0$, which gives $M_p=M_n=938.9$ MeV and $M_\Lambda=1113.7$ MeV by applying Eq.~(\ref{eq:Bmass}). Meanwhile, as was done in traditional RMF models~\cite{Meng2016_RDFNS}, in addition to the repulsive interaction exerted by vector mesons, we take $\alpha_S = 0.57$ at $n_\mathrm{b}=n_S=0.16\ \mathrm{fm}^{-3}$ so that a strong attractive interaction ($M_p=M_n=519.2$ MeV) is present to accommodate the spin-orbit splittings in finite nuclei, which is essential to reproduce the nuclear magic numbers. We then write out the following formula
\begin{equation}
  \alpha_S = a_S \exp(-n_\mathrm{b}/n_S)+b_S,  \label{eq:alphaS}
\end{equation}
which reproduce the values $\alpha_S(0) = 0.849$ and $\alpha_S(n_S) = 0.57$ with $a_S$, $n_S$ and $b_S$ presented in Table \ref{table:DDparam}. In Fig.~\ref{Fig:Mi_sym} we present the masses of nucleons, $\Lambda$-hyperons, and quarks in symmetric nuclear matter (SNM), which are obtained with Eqs.~(\ref{eq:Bmass}) and (\ref{eq:qmass}). Evidently, the masses of nucleons and $\Lambda$-hyperons decrease quickly at $n_\mathrm{b}\lesssim n_S$ due to the structural function~(\ref{eq:alphaS}). At larger densities, the masses of $u$-quarks, $d$-quarks, and nucleons further decrease and approach to their current masses with vanishing quark condensates, while the $s$-quark condensate $\bar{n}_{s}^{s}$ remains large at $n_\mathrm{b}\lesssim 1\ \mathrm{fm}^{-3}$ so that the masses of $s$-quarks and $\Lambda$-hyperons are still sizable in Fig.~\ref{Fig:Mi_sym}.

\begin{table}
  \centering
  \caption{\label{table:DDparam} The adopted parameter sets for the density dependent coupling constants in Eqs.~(\ref{eq:alphaS}-\ref{eq:alphaTV}).}
  \begin{tabular}{l|l|l}
    \hline \hline
  $a_S=0.4413715$     & $n_S=0.16\ \mathrm{fm}^{-3}$    & $b_S=0.4076285$ \\
  $a_V=3.566049$     & $n_V=0.214\ \mathrm{fm}^{-3}$    & $b_V=1.062771$  \\
  $a_{TV}=0.5014459$  & $n_{TV}=0.1\ \mathrm{fm}^{-3}$  & $b_{TV}=0.0117601$ \\     \hline
  \end{tabular}
\end{table}

\begin{figure}[!ht]
  \centering
  \includegraphics[width=0.95\linewidth]{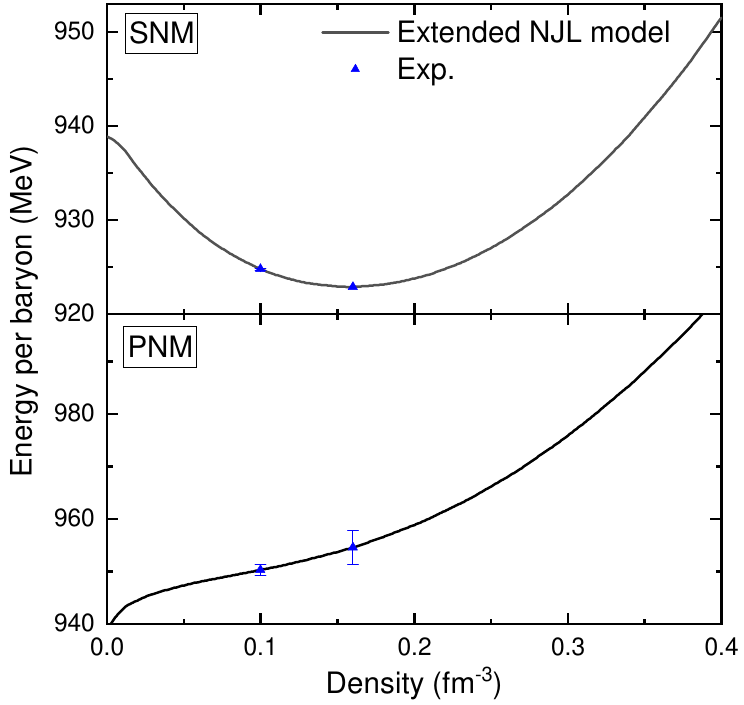}
  \caption{\label{Fig:EpA}Energy per baryon $E/n_\mathrm{b}$ of nuclear matter as functions of the total baryon number density $n_\mathrm{b}$, which are obtained adopting the parameter sets indicated in Table.~\ref{table:DDparam}.}
\end{figure}

To fix the coupling constants of vector mesons, we adopt the well-constrained nuclear matter properties at $n_0 = 0.16 \ \mathrm{fm}^{-3}$ and $n_\mathrm{on} = 0.1 \ \mathrm{fm}^{-3}$, i.e., $\varepsilon(n_0) = -16$ MeV, $S(n_0) = 31.7 \pm 3.2$ MeV~\cite{Li2013_PLB727-276, Oertel2017_RMP89-015007}, $\varepsilon(n_\mathrm{on}) = -14.1\pm0.1$ MeV, and $S(n_\mathrm{on})=25.5\pm1.0$ MeV~\cite{Centelles2009_PRL102-122502, Brown2013_PRL111-232502}. In practice, with the nucleon masses being $m_N=M_p(0)=M_n(0)=938.9$ MeV, we reproduce the energy per baryon $E(n_\mathrm{on})/n_\mathrm{on} =m_N+\varepsilon(n_\mathrm{on})= 924.8$ MeV and $E(n_0)/n_0 = m_N+\varepsilon(n_0)=922.9$ MeV for SNM and $E(n_\mathrm{on})/n_\mathrm{on} =m_N+\varepsilon(n_\mathrm{on})+S(n_\mathrm{on})= 950.3$ MeV and $E(n_0)/n_0 = m_N+\varepsilon(n_0)+S(n_0)=954.6$ MeV for pure neutron matter (PNM) adopting the following density dependent coupling constants, i.e.,
\begin{eqnarray}
g_\omega^2/m_\omega^2 &=& 4G_S [a_V\exp(-n_\mathrm{b}/n_V) + b_V], \label{eq:alphaV}\\
g_\rho^2/m_\rho^2 &=& 4G_S [a_{TV}\exp(-n_\mathrm{b}/n_{TV}) + b_{TV}]. \label{eq:alphaTV}
\end{eqnarray}
The coefficients in Eqs.~(\ref{eq:alphaV}) and (\ref{eq:alphaTV}) used in this work are indicated in Table \ref{table:DDparam}. Note that the meson masses $m_{\sigma}$, $m_{\omega}$, and $m_{\rho}$ are left undetermined in this work, which should be fixed in our future study according to the properties of finite nuclei. Currently, they only appear in combination with the couplings, i.e.,  $g_{\sigma}^2/m_{\sigma}^2=4G_S$, and $g_{\omega,\rho}^2/m_{\omega,\rho}^2$ as indicated in Eqs.~(\ref{eq:alphaV}) and (\ref{eq:alphaTV}).

\begin{table}
\caption{\label{table:NM} The saturation properties of nuclear matter predicted by the extended NJL model introduced in this work. }
\begin{tabular}{c|cccc} \hline \hline
 $n_0$        &   $\varepsilon$    &   $K$  &  $S$   & $L$      \\
 fm${}^{-3}$  &   MeV              &   MeV  &   MeV  &  MeV     \\   \hline
  0.158       &    $-16.0$         & 234.5 &  31.5  & 42.4      \\
\hline
\end{tabular}
\end{table}

In Fig.~\ref{Fig:EpA} we then present the energy per baryon of SNM and PNM, which are obtained with Eq.~(\ref{eq:ener}) adopting the parameter sets indicated in Table.~\ref{table:DDparam}. Evidently, the binding energies of both SNM and PNM reproduce the central values of the binding energies $\varepsilon(n_0) = -16$ MeV, $\varepsilon(n_\mathrm{on}) = -14.1\pm0.1$ MeV and symmetry energies $S(n_0) = 31.7 \pm 3.2$ MeV,  $S(n_\mathrm{on})=25.5\pm1.0$ MeV~\cite{Li2013_PLB727-276, Oertel2017_RMP89-015007, Centelles2009_PRL102-122502, Brown2013_PRL111-232502}. The corresponding saturation properties such as the incompressibility $K$ and slope $L$ of nuclear symmetry energy are indicated in Table \ref{table:NM}, which are consistent with the state-of-art constraints $K = 240 \pm 20$ MeV~\cite{Shlomo2006_EPJA30-23} and $L = 58.7 \pm 28.1$ MeV~\cite{Li2013_PLB727-276, Oertel2017_RMP89-015007}. Note that constraints on higher order terms of nuclear saturation properties can be attained based astrophysical observations, heavy-ion collisions, measurements of the neutron skin thicknesses, and nuclear theories~\cite{Zhang2020_PRC101-034303, Xie2021_JPG48-025110, PREX2021_PRL126-172502, Essick2021_PRL127-192701, CREX2022_PRL129-042501}.

\begin{figure}[!ht]
  \centering
  \includegraphics[width=0.95\linewidth]{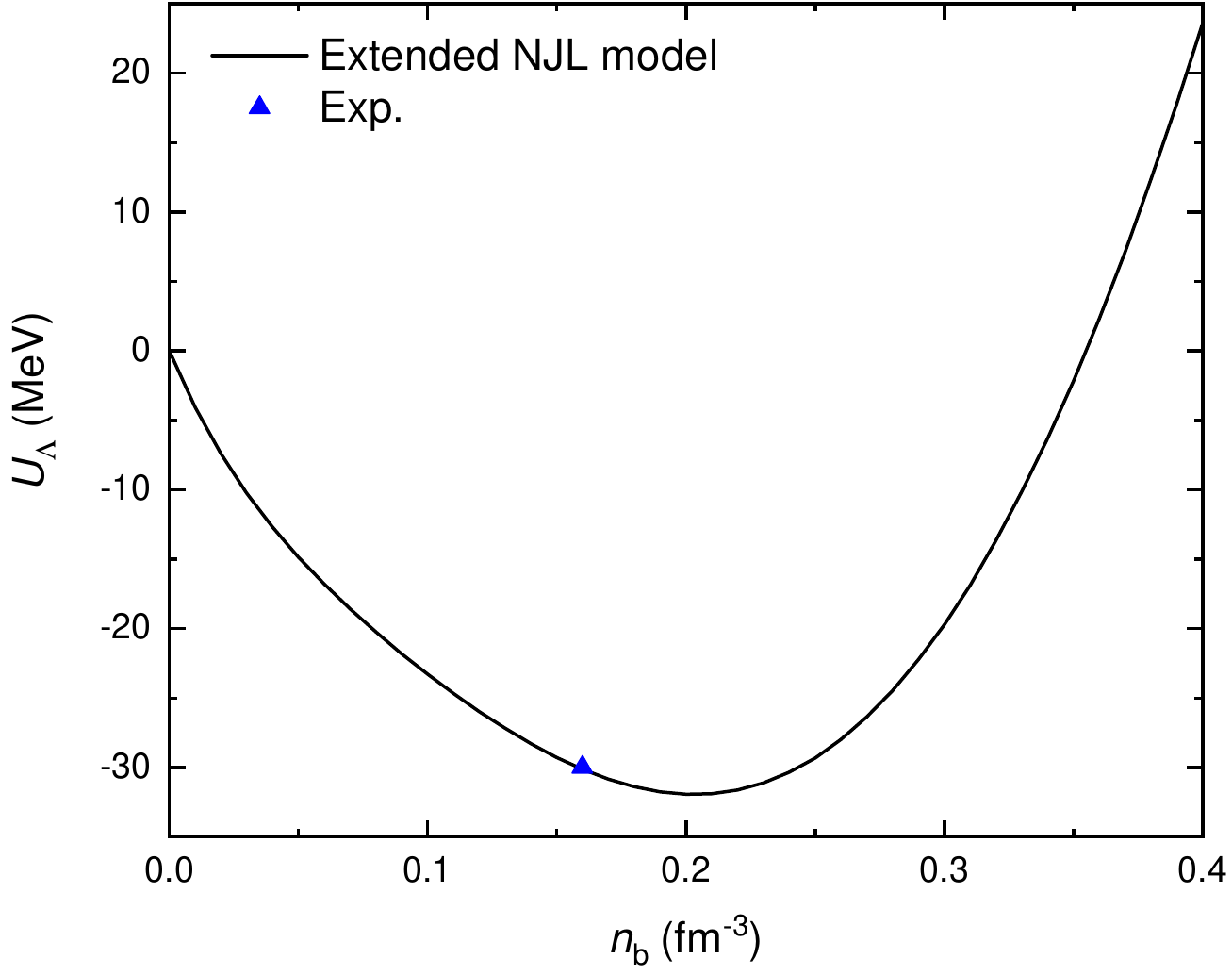}
  \caption{\label{Fig:ULambda} Potential depth of $\Lambda$-hyperons in symmetric nuclear matter.}
\end{figure}

Beside nucleons, the contribution of $\Lambda$-hyperons is also considered in this work. Based on previous investigations of $\Lambda$-hypernuclei, it was shown that a $\Lambda$ potential well depth $U_\Lambda(n_0) = - 30$~MeV in SNM is required to accommodate the single $\Lambda$ binding energies~\cite{Gal2016_RMP88-035004, Sun2016_PRC94-064319, Sun2018_CPC42-25101, Rong2021_PRC104-054321}. In the framework of the extended NJL model, the potential depth $U_\Lambda(n_0) = - 30$~MeV indicates $f_\Lambda =1.0626$ for the vector meson couplings. The corresponding potential depth of $\Lambda$-hyperons is then fixed by $U_\Lambda(n_\mathrm{b}) = \mu_\Lambda - M_\Lambda(0)$ with $\nu_\Lambda=0$, where the corresponding values are presented in Fig.~\ref{Fig:ULambda}. The onset densities of $\Lambda$-hyperons in SNM and PNM are then $n_\mathrm{onset} = 0.72$ and $0.44 \ \mathrm{fm}^{-3}$, respectively.

\section{\label{sec:res}Results and discussions}
Based on the formulae and model parameters presented in Sec.~\ref{sec:the}, we then investigate the properties of dense stellar matter, where baryonic matter, quark matter, and quarkyonic matter can be treat in a unified manner. For compact star matter at a fixed total baryon number density $n_\mathrm{b}$, the densities of fermions are obtained with Eqs.~(\ref{eq:chem_b}-\ref{eq:chem_l}) fulfilling the $\beta$-stability condition, i.e.,
\begin{equation}
\mu_i= B_i \mu_\mathrm{b} - q_i \mu_e,  \label{eq:weakequi}
\end{equation}
where $\mu_\mathrm{b}$ is the baryon chemical potential with $B_i$ ($B_p=B_n=B_\Lambda=1$, $B_u=B_d=B_s=1/3$, and $B_e=B_\mu=0$) being the baryon number of particle type $i$. Due to charge screening, the charge neutrality condition needs to be satisfied as well, i.e.,
\begin{equation}
  \sum_{i} q_i n_i = 0,
\end{equation}
where $q_{n}=0$, $q_{p}=1$, $q_{u}=2/3$, $q_{d}=-1/3$, and $q_{e}=q_{\mu}=-1$ are the charge number of each particle type. The EOSs of compact star matter are obtained with the energy density $E$ fixed by Eq.~(\ref{eq:ener}) and pressure $P$ by Eq.~(\ref{eq:pressure}).

\begin{figure}[!ht]
  \centering
  \includegraphics[width=0.95\linewidth]{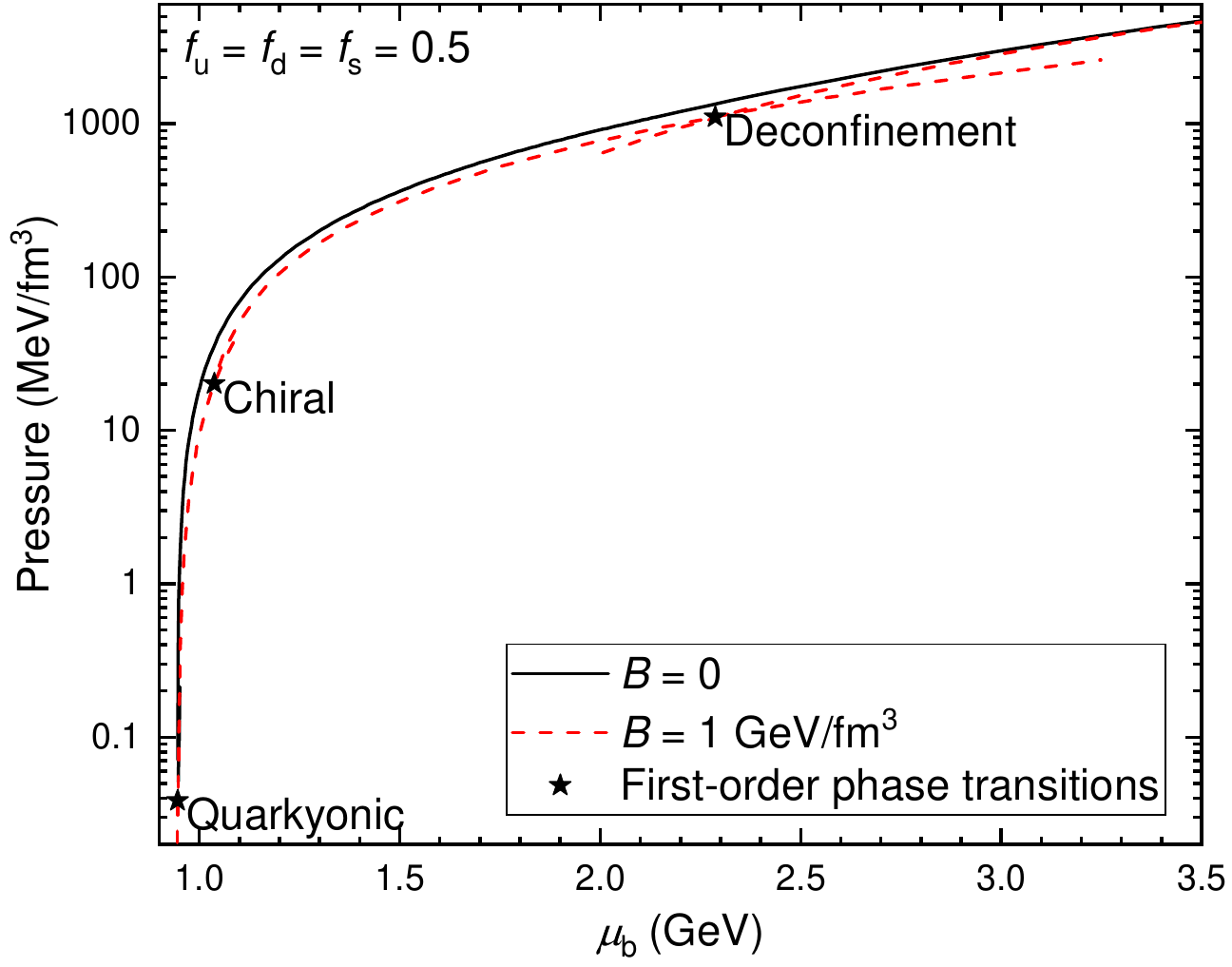}
  \caption{\label{Fig:Pressure_mun} Pressure $P$ as functions of baryon chemical potential $\mu_\mathrm{b}$ adopting various parameter sets with $f_u=f_d=f_s=0.5$, $B=0$ and 1 GeV/fm$^3$. A first-order quarkyonic transition is observed if we take $B=0$, while first-order chiral and deconfinement transitions are identified if $B=1$ GeV/fm$^3$.}
\end{figure}

We first examine the phase structure of dense stellar matter taking $f_u=f_d=f_s=0.5$, which approximately corresponds to the vector coupling $G_V\approx f_q b_VG_S = 0.55 G_S$ in traditional NJL models. Two scenarios for the Pauli blocking term in Eq.~(\ref{eq:Bmass}) are considered, i.e., $B=0$ and 1 GeV/fm$^3$. For uniform compact star matter, the thermodynamical potential density is $\Omega=-P$. The stable phase at a fixed baryon chemical potential $\mu_\mathrm{b}$ is fixed by minimizing the thermodynamical potential density $\Omega$ or maximizing the pressure $P$.

\begin{figure*}[!ht]
  \centering
  \includegraphics[width=0.75\linewidth]{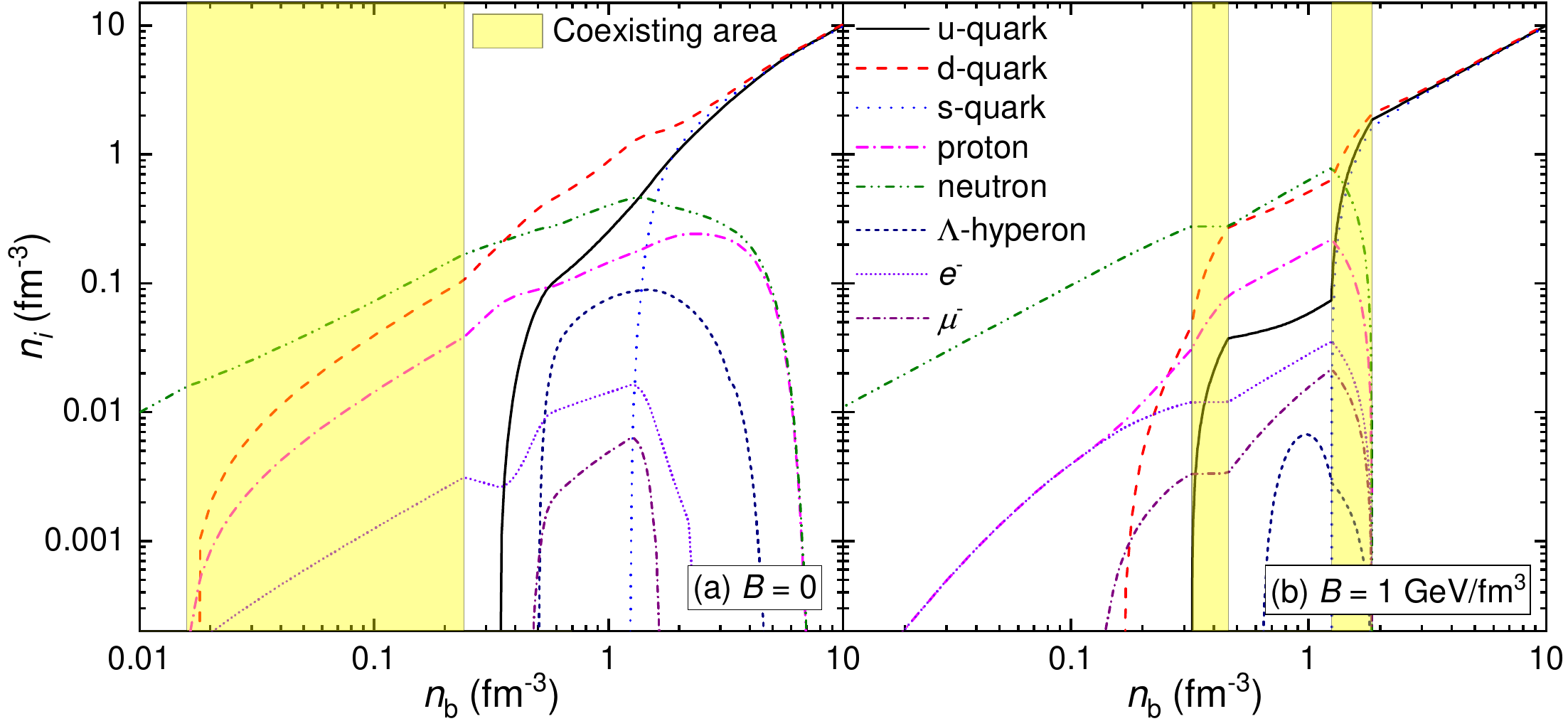}
  \caption{\label{Fig:ni_fV05} Number densities of various fermions considered in this work. The yellow bands correspond to the density ranges of mixed phases indicated by the solid stars in Fig.~\ref{Fig:Pressure_mun}, which are fixed by Maxwell construction.  }
\end{figure*}

In Fig.~\ref{Fig:Pressure_mun} we present the obtained pressure as functions of baryon chemical potential. Various first-order phase transitions are identified, where the solid stars indicate the coexistence points of those first-order phase transitions according to Maxwell construction of mixed phases. In principle, the Maxwell construction is valid only when the surface tension is large enough so that geometrical structures are unfavorable, which is often not the case~\cite{Heiselberg1993_PRL70-1355, Voskresensky2002_PLB541-93, Tatsumi2003_NPA718-359, Voskresensky2003_NPA723-291, Endo2005_NPA749-333, Maruyama2007_PRD76-123015, Yasutake2014_PRC89-065803, Xia2019_PRD99-103017, Maslov2019_PRC100-025802, Xia2020_PRD102-023031}. A more detailed investigation on the structures and properties of the mixed phases is thus necessary and should be carried out in our future study, while for now we restrict ourself to the Maxwell construction scenarios. In this work, we consider the following phase transitions:
\begin{itemize}
  \item Quarkyonic phase transition: emergence of quasi-free quarks in baryonic matter, where quarks are still confined and only baryons can be exited to higher energy states;
  \item Chiral phase transition: quark condensate $\bar{n}_{q}^{s}$ ($q=u,d,s$) vanishes and quark mass $M_q$ is reduced to its current mass $m_{q0}$;
  \item Deconfinement phase transition: at large enough densities baryons become unbound and start to dissolve, i.e., Mott transition.
\end{itemize}
All three types of phase transitions are identified in dense stellar matter, which become of first order if we adopt certain parameter sets. For example, as indicated in Fig.~\ref{Fig:Pressure_mun}, the quarkyonic phase transition is of first order if we take $B=0$, which becomes of third order for $B=1$ GeV/fm$^3$. Meanwhile, the chiral and deconfinement phase transitions become of first order only when we take large $B$, e.g., $B=1$ GeV/fm$^3$.

The quarkyonic and deconfinement phase transitions are better illustrated in Fig.~\ref{Fig:ni_fV05}, where the number densities of various fermions are plotted. In particular, due to the requirement of charge neutrality, $d$-quarks emerge at $n_\mathrm{b} = n_0/2$ if we take $B=0$, which leads to a first-order phase transition with the reduction of energy per baryon. In such cases, the corresponding uniform baryonic matter and quarkyonic matter become unstable, then the onset density for $d$-quarks is decreased due to the coexistence of the baryonic matter ($n_\mathrm{b} \approx 0.016$ fm$^{-3}$) and quarkyonic matter ($n_\mathrm{b} \approx $ 0.24 fm$^{-3}$). A mixed phase at $n_\mathrm{b} \approx 0.016$-0.24 fm$^{-3}$ is then predicted by Maxwell construction, corresponding to the bulk separation of the quarkyonic and baryonic phases. Such an early emergence of quarkyonic phase is thus unfavorable according to the nuclear saturation properties. In fact, as indicated in Fig.~\ref{Fig:MR}, such an early onset of quarkyonic phase predicts too small radii for neutron stars, which is in strong tension with various astrophysical observations. If we adopt $B=1$ GeV/fm$^3$, due to strong repulsive interactions between baryons and quarks, $d$-quarks emerge at much larger densities at $n_\mathrm{b} = 0.17$ fm$^{-3}$, while the quarkyonic phase transition becomes of third order. Similar scenarios are observed if we adopt larger $f_q$.

Meanwhile, it is found that the onset densities of $\Lambda$-hyperons are much larger with $n_\mathrm{onset} = 0.51$ and 0.64 fm$^{-3}$ when we take $B=0$ and 1 GeV/fm$^3$, respectively. Due to the strong $\Lambda$-$\omega$ coupling, $\Lambda$-hyperons eventually vanish at larger densities. In such cases, $\Lambda$-hyperons have little impacts on the EOSs of compact star matter, so that the Hyperon Puzzle can be avoided~\cite{Vidana2015_AIPCP1645-79}.

As we further increase the density, baryons eventually become unbound and the corresponding number densities decrease quickly. A deconfinement phase transition then takes place as baryons vanish at $n_\mathrm{b} \approx 7$ (1.3) fm$^{-3}$ when we take $B=0$ (1) GeV/fm$^3$, which is of third (first) order. At largest densities, as indicated in Fig.~\ref{Fig:ni_fV05}, the quark number densities in strange quark matter become approximately the same with $n_u=n_d=n_s$, which favors the formation of color-flavor locked phase~\cite{Ruester2005_PRD72-034004}. This is nevertheless out of the scope of our current study and should be examined in future. Note that the density for deconfinement phase transition increases significantly as $f_q$ ($q=u,d,s$) approaches to 1.

\begin{figure*}[!ht]
  \centering
  \includegraphics[width=0.75\linewidth]{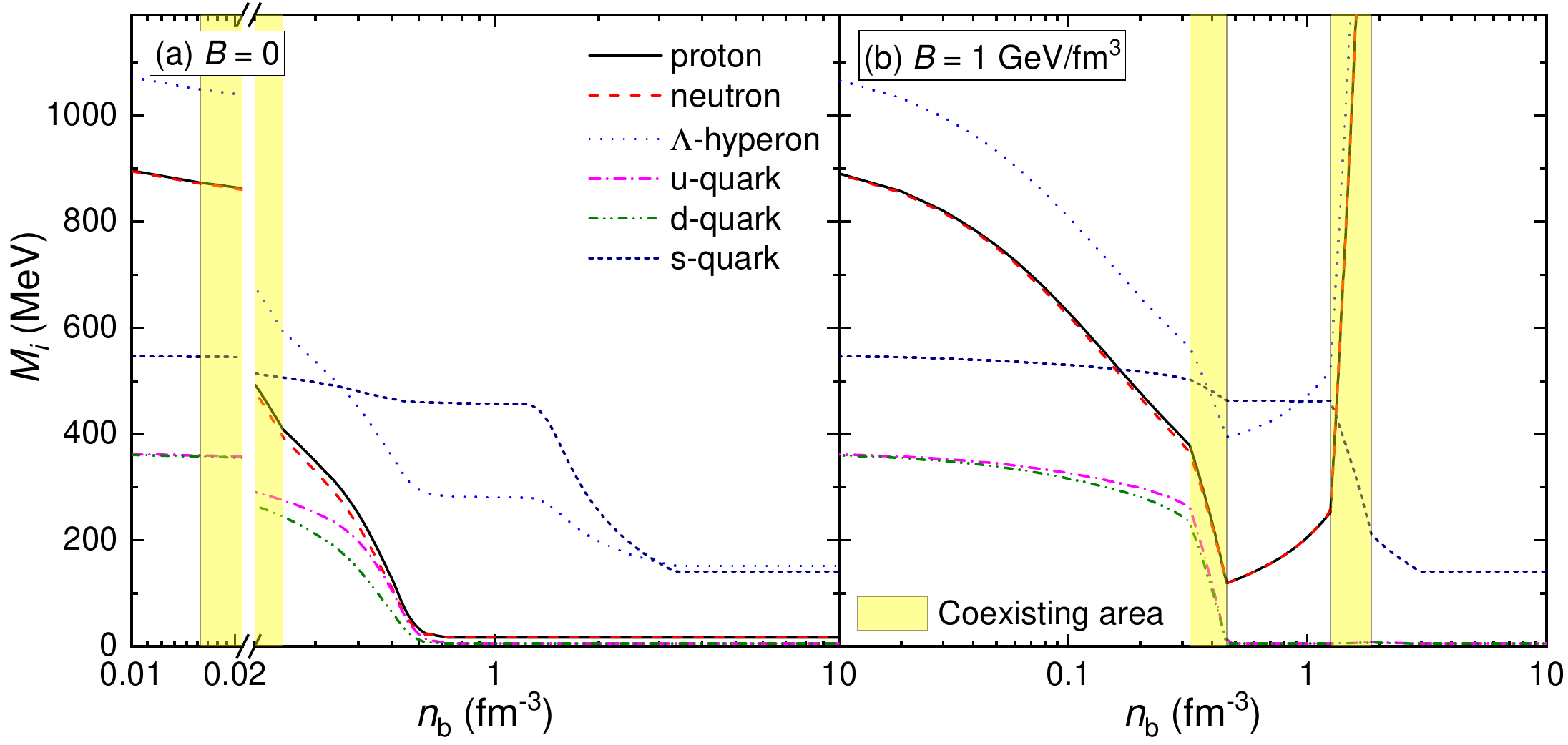}
  \caption{\label{Fig:Mi_fV05} Masses of nucleons, $\Lambda$-hyperons, and quarks in dense stellar matter as functions of the total baryon number density $n_\mathrm{b}$, corresponding to those in Figs.~\ref{Fig:Pressure_mun} and~\ref{Fig:ni_fV05}.}
\end{figure*}

In between the quarkyonic and deconfinement phase transitions, a chiral phase transition takes place, where the quark condensates $\bar{n}_{u}^{s}$ and $\bar{n}_{d}^{s}$ vanish. It is found that the chiral phase transition is usually a smooth crossover when we take $B=0$, which becomes of first order if $B=1$ GeV/fm$^3$. As quark condensates determine the masses of baryons and quarks based on Eqs.~(\ref{eq:Bmass}) and (\ref{eq:qmass}), in Fig.~\ref{Fig:Mi_fV05} we then present their masses as order parameters for chiral phase transitions. Evidently, after quarkyonic phase transition takes place, the masses of $u$- and $d$-quarks decrease quickly with density and approach to their current masses, indicating the chiral phase transition with vanishing $\bar{n}_{u}^{s}$ and $\bar{n}_{d}^{s}$. Note that in this phase, there are still vector interactions from baryon as indicated in Eq.~(\ref{eq:Sigma_Q}), while Pauli blocking term in Eq.~(\ref{eq:Bmass}) also plays a role.

The baryon masses decrease with density as well, where the mass of $\Lambda$-hyperons becomes even smaller than $s$-quarks. This is mainly attributed to the structural function $\alpha_S$ ($<1$) adopted in this work, where three quarks form a bound state with large binding energies. The mass of $s$-quarks will decrease at much larger densities with vanishing $s$-quark condensate $\bar{n}_{s}^{s}$, which may be accompanied with a first-order deconfinement phase transition if we take $B=1$ GeV/fm$^3$. The effects of Pauli blocking term in Eq.~(\ref{eq:Bmass}) become evident at large densities, where the masses of baryons increase drastically with $n_\mathrm{b}^Q$ if we adopt $B=1$ GeV/fm$^3$. Consequently, as indicated in Fig.~\ref{Fig:ni_fV05}, baryons begin to dissolve with the corresponding number densities approaching zero.

\begin{figure}
\includegraphics[width=0.9\linewidth]{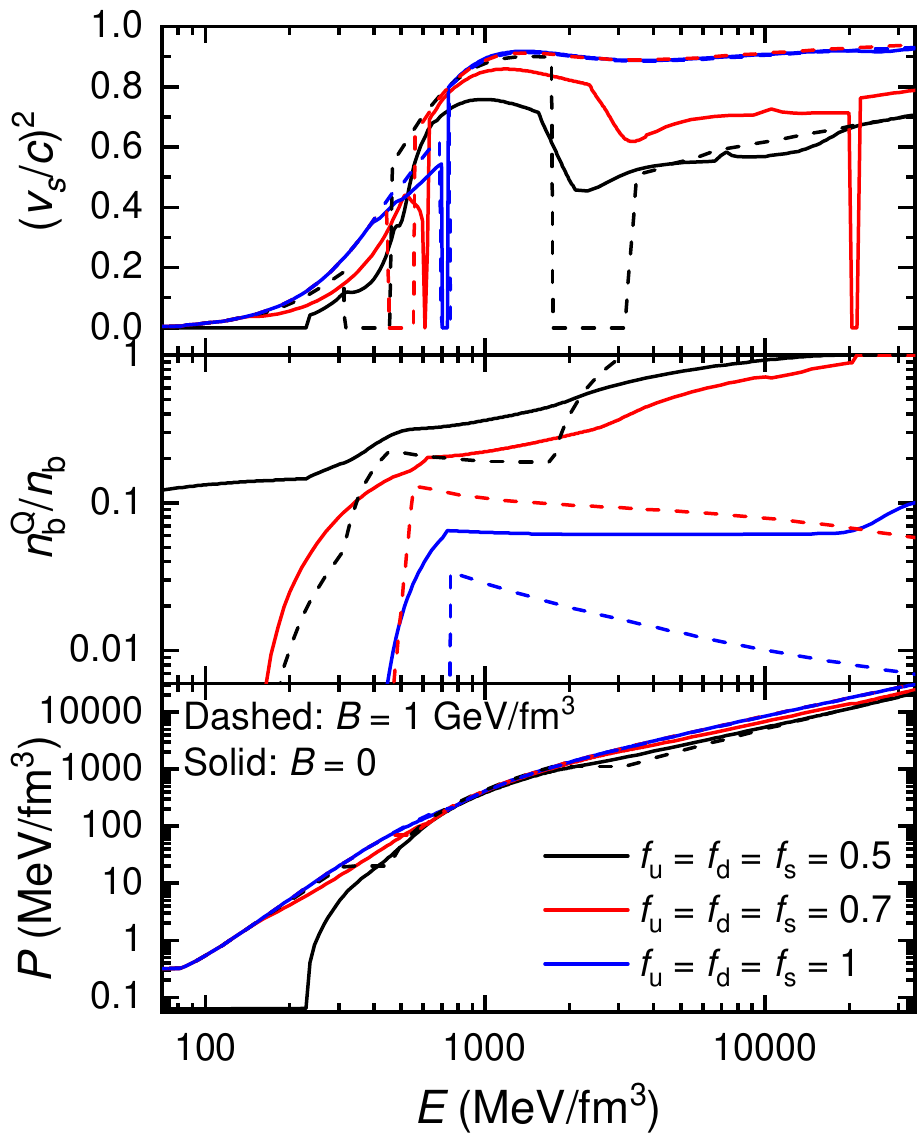}
\caption{\label{Fig:EOS} Velocity of sound $v_s$ (upper panel), quark fraction $n_\mathrm{b}^Q/n_\mathrm{b}$ (middle panel) and pressure $P$ (lower panel) of compact star matter as functions of energy density, which are obtained adopting the parameters $f_u=f_d=f_s=0.5$ (black), 0.7 (red), 1 (blue) and $B=0$, 1 GeV/fm$^3$.}
\end{figure}

In Fig.~\ref{Fig:EOS}, we present the velocity of sound $v_s=\sqrt{\mbox{d}P/\mbox{d}E}$, quark fraction $n_\mathrm{b}^Q/n_\mathrm{b}$, and the corresponding EOSs predicted by the extended NJL model, where various combinations of parameters $f_q=0.5$, 0.7, 1 and $B=0$, 1 GeV/fm$^3$ are adopted. At $P\lesssim 1$ MeV/fm$^3$, a first-order core-crust phase transition takes place. Since the saturation properties of nuclear matter indicated in Table~\ref{table:NM} resemble those of DD-LZ1 in RMF models, we adopt the DD-LZ1 EOS for neutron star crusts~\cite{Xia2022_CTP74-095303, Xia2022_PRC105-045803}, where the core-crust transition pressure is fixed by Maxwell construction. This nevertheless should be improved in our future study, where the EOSs and structures for neutron star crusts should be fixed in the extended NJL model in a unified manner. At larger pressures, quarkyonic phase transition takes place with nonzero $n_\mathrm{b}^Q$, which is generally of third order except for the case with a first-order quarkyonic phase transition obtained by adopting $f_q=0.5$ and $B=0$. As density increases, quark condensates $\bar{n}_{u}^{s}$ and $\bar{n}_{d}^{s}$ eventually approach to zero at $P\gtrsim 10$ MeV/fm$^3$, indicating the occurrence of chiral phase transitions. Most chiral phase transitions are of first order, which becomes a smooth crossover if we take $f_q=0.5$ and $B=0$ with a large quark fraction. As indicated in the upper and middle panels of Fig.~\ref{Fig:EOS}, finally a deconfiement phase transition takes place with $n_\mathrm{b}^Q/n_\mathrm{b}=1$ if we adopt the parameter sets ($f_q$, $B$ in GeV/fm$^3$): (0.5, 0), (0.5, 1), and (0.7, 0), where the EOSs are dominated by quark matter with $n_u\approx n_d\approx n_s$ and velocity of sound reduced. Meanwhile, within the density range indicated in Fig.~\ref{Fig:EOS}, baryonic matter still dominates if we adopt the parameter sets (0.7, 1), (1, 0), and (1, 1), where the EOSs coincide with each other at large densities with a large velocity of sound.

\begin{figure}
  \centering
  \includegraphics[width=\linewidth]{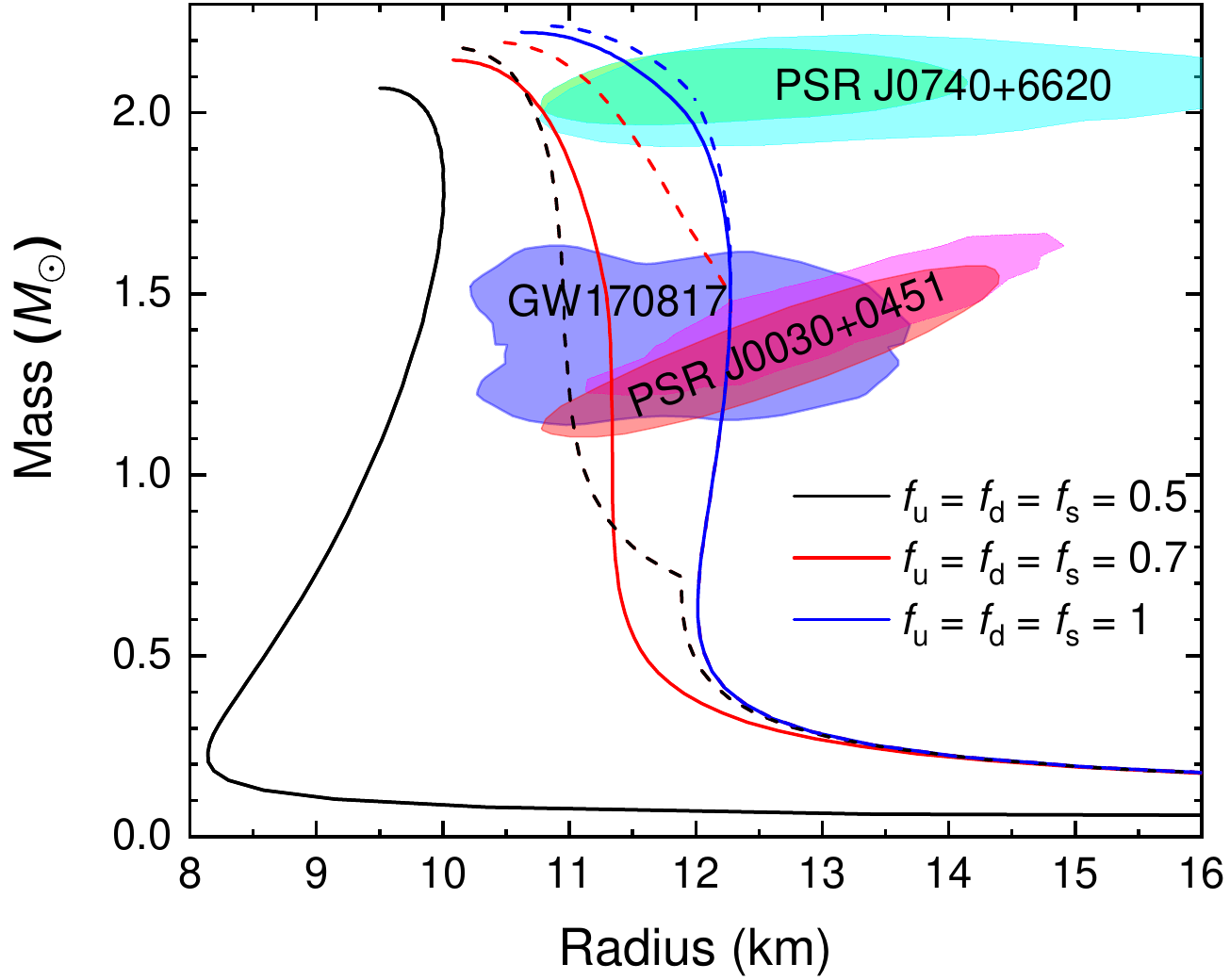}
\caption{\label{Fig:MR} Mass-radius relations of compact stars obtained with the EOSs presented in Fig.~\ref{Fig:EOS}. The shaded regions indicate the mass-radius constraints from the binary neutron star merger event GW170817~\cite{LVC2018_PRL121-161101}, PSR J0030+0451 and PSR J0740+6620~\cite{Riley2019_ApJ887-L21, Riley2021_ApJ918-L27, Miller2019_ApJ887-L24, Miller2021_ApJ918-L28}.}
 \label{Fig:MRall}
\end{figure}

Based on the EOSs presented in Fig.~\ref{Fig:EOS}, we then examine the corresponding compact star structures by solving the Tolman-Oppenheimer-Volkov (TOV) equation
\begin{equation}
\frac{\mbox{d}P}{\mbox{d}r} = \frac{(E+P)(M+4\pi r^3 P)} {2 M r-r^2/G}\ \text{and } \frac{\mbox{d}M}{\mbox{d}r} = 4\pi E r^2   \label{eq:TOV}
\end{equation}
with the gravity constant $G=6.707\times 10^{-45}\ \mathrm{MeV}^{-2}$. The obtained $M$-$R$ relations of compact stars are then presented in Fig.~\ref{Fig:MR}. Various observational constraints on compact star structures are indicated by the shaded areas, i.e., the masses and radii fixed by pulse profile modeling of PSR J0030+0451 and PSR J0740+6620~\cite{Riley2019_ApJ887-L21, Riley2021_ApJ918-L27, Miller2019_ApJ887-L24, Miller2021_ApJ918-L28} as well as the gravitational wave measurements from of the binary neutron star merger event GW170817~\cite{LVC2018_PRL121-161101}. Additionally, there are precise mass measurements for PSR J1614-2230 (1.928 $\pm$ 0.017 $M_{\odot}$)~\cite{Fonseca2016_ApJ832-167} and PSR J0348+0432 (2.01 $\pm$ 0.04 $M_{\odot}$)~\cite{Antoniadis2013_Science340-1233232} by analyzing the orbital motion of pulsars in a binary system~\cite{Lattimer2012_ARNPS62-485}, which is not indicated in Fig.~\ref{Fig:MR} since PSR J0740+6620 also reaches the two-solar-mass constraint.

The $M$-$R$ relations of compact stars predicted by the extended NJL model adopting the parameter sets (0.7, 1), (1, 0), and (1, 1) agree well with the observational constraints, while the radii for two-solar-mass compact stars predicted by the parameter sets (0.5, 1) and (0.7, 0) lie in the lower ends of the PSR J0740+6620 constraints~\cite{Riley2021_ApJ918-L27, Miller2021_ApJ918-L28}. Additionally, if the parameter set (0.5, 0) is adopted, the radii of the corresponding compact stars become too small according to the observational constraints, suggesting that the first-order quarkyonic phase transition at small densities is in tension with pulsar observations.

\section{\label{sec:con}Conclusion}

In this work, we propose an extended NJL model to describe baryonic matter, quark matter, and their transitions in a unified manner. The baryons are treated as clusters made of three quarks, while a density-dependent structural function $\alpha_S$ is introduced to modulate the four(six)-point interaction strengths to reproduce the baryon masses in vacuum and medium~\cite{Bentz2001_NPA696-138, Reinhardt2012_PRD85_074029, Xia2014_CPL31-41101}. Additional vector interactions are introduced using vector mesons $\omega$ and $\rho$, where the baryon-meson couplings are density-dependent and are fixed by reproducing nuclear matter properties~\cite{Li2013_PLB727-276, Oertel2017_RMP89-015007, Centelles2009_PRL102-122502, Brown2013_PRL111-232502} and $\Lambda$-hyperon potential depth in nuclear medium~\cite{Gal2016_RMP88-035004, Sun2016_PRC94-064319, Sun2018_CPC42-25101, Rong2021_PRC104-054321}. A dampening factor $f_q\equiv f_u=f_d=f_s$ is introduced to modulates the coupling strengths between quarks and vector mesons, where in this work we take $f_q=0.5$, 0.7, and 1. Similar to the melting of light clusters in nuclear medium~\cite{Typel2010_PRC81-015803}, we have introduced a Pauli blocking term to baryon masses so that baryons eventually become unbound in the presence of a quark Fermi Sea~\cite{Xia2018_JPSCP20-011010, Xia2023_PRD108-054013}. Then the deconfinement phase transition can be viewed as a Mott transition of quark clusters~\cite{Bastian2018_Universe4-67}.

Employing the extended NJL model, we then investigate the properties of dense stellar matter. As the density of nuclear matter increases, quarks emerge as quasi-free particles and coexist with baryons in the same volume. This phase is considered as quarkyonic matter, where quarks can not be exited to higher energy states in the presence of baryons due to confinement~\cite{McLerran2007_NPA796-83, McLerran2009_NPA830-709c, McLerran2019_PRL122-122701}. Depending on the strengths of Pauli blocking term $B$ and quark-vector meson couplings $f_q$, both first-order and continues phase transitions are observed for quarkyonic, chiral, and deconfinement phase transitions. In particular, we find quarkyonic phase transition is generally of third order but becomes of first order if we adopt $f_q=0.5$ and $B=0$, which predicts compact stars that are too compact according to observational constraints. Similar situation takes place for chiral phase transitions, which are generally of first order but becomes a smooth crossover if we take $f_q=0.5$ and $B=0$. At largest densities, a deconfinement phase transition takes place with baryons melted, where both first-order and continues transitions are identified. Adopting Maxwell construction for the first-order phase transitions, we then obtain the EOSs of dense stellar matter. The corresponding compact star structures are then fixed by solving the TOV equation, which are confronted with various astrophysical constraints~\cite{LVC2018_PRL121-161101, Riley2019_ApJ887-L21, Riley2021_ApJ918-L27, Miller2019_ApJ887-L24, Miller2021_ApJ918-L28}. Our prediction generally agrees well with observations, while the parameter set $f_q=0.5$ and $B=0$ predicts too small radii for compact stars.

\begin{acknowledgments}
The author would like to thank Professors Kenji Fukushima, Bao-An Li, Toshiki Maruyama, Makoto Oka, Gubler Philipp, Ting-Ting Sun, Toshitaka Tatsumi and Nobutoshi Yasutake for fruitful discussions. This work was supported by the National Natural Science Foundation of China (Grant No. 12275234) and the National SKA Program of China (Grant No. 2020SKA0120300).
\end{acknowledgments}


%

\end{document}